\documentclass[intlimits,twoside,a4paper]{article}

\usepackage{amsmath,amssymb}
\usepackage{graphicx}
\usepackage{siunitx}

\usepackage[T2A]{fontenc}
\usepackage[cp1251]{inputenc}

\usepackage{cmpj2}

\articletype{Proceedings Paper}

\issue{2013}{16}{3}{31705}
\doinumber{10.5488/CMP.16.31705}

\title[An attempt to obtain Bi$_4$Ti$_3$O$_{12}$-PVC textured ceramics-polymer composites]%
{An attempt to obtain Bi$_4$Ti$_3$O$_{12}$-PVC textured ceramics-polymer composites}
\author[E. Nogas-\'{C}wikiel, H. Bernard]{E. Nogas-\'{C}wikiel\refaddr{label1}\thanks{E-mail: ewa.nogas-cwikiel@us.edu.pl} \ ,
        H. Bernard\refaddr{label2}}
\addresses{
\addr{label1} Faculty of Computer Science and Materials Science, University of Silesia, Sosnowiec, Poland
\addr{label2} Scholar of the project ``DoktoRIS'', Poland
}

\date{Received October 23, 2012}
\authorcopyright{E. Nogas-\'{C}wikiel, H. Bernard, 2013}

\begin{document}

\maketitle

\begin{abstract}
Bi$_4$Ti$_3$O$_{12}$-PVC composites were fabricated. Ceramics powders of bismuth titanate were prepared by the sol-gel method using bismuth nitrate pentahydrate Bi(NO$_3$)$_3$\,$\cdot$\,5H$_2$O and tetrabutyl titanate Ti(CH$_3$(CH$_2$)$_3$O)$_4$ as precursors. The Bi$_4$Ti$_3$O$_{12}$-PVC composites were fabricated from ceramics powders and polymer powders by hot-pressing method.
\pacs 77.84.-s, 77.84.Lf, 77.84.Dy, 81.20.Fw
\end{abstract}

\section{Introduction}

Scientists and practitioners working in piezoelectronics, electroacoustics, hydroacoustics, optoelectronics and in other fields of modern technology that use sensors and electroceramic transducers aim at obtaining such materials to produce transducers, which give as high response as possible. It is difficult to gain high piezoelectric responses with randomly oriented polycrystalline ceramic materials. It was recently proved, on several types of ferroelectric and piezoelectric materials such as PMN-PT \cite{1}, Ba$_6$Ti$_{17}$O$_{40}$-PZT \cite{2}, SrFe$_{12}$O$_{19}$ \cite{3} and Bi$_4$Ti$_3$O$_{12}$ \cite{4}, that their properties could be optimized thanks to a preferential crystalline orientation.
	
Monolithic ceramics are dense, brittle and are of a limited size and shape. These defects may be eliminated by using ceramics-polymer composites instead of monolithic ceramics for the construction of transducers. Composites with a polymer base and ceramic powder as the active phase are better than monolithic ceramics in that they may be used in any shape and size, they are lighter and have better mechanical resistance.

In the work, an attempt was made to obtain Bi$_4$Ti$_3$O$_{12}$-PVC textured ceramics-polymer composites.

\section{Choice of material and its synthesis method}

It became possible to obtain a textured composite thanks to:
\begin{itemize}
  \item the proper choice of ceramics,
  \item the sol-gel method used for its synthesis.
\end{itemize}

It was assumed that it is possible to obtain the texture for materials that grow anisotropically. Ceramic materials with asymmetric unit cells or crystalline structure composed of chains or polyhedral layers often grow anisotropically.

\begin{figure}[!t]
\begin{center}
\begin{tabular}{cc}
\includegraphics[width=0.3\textwidth]{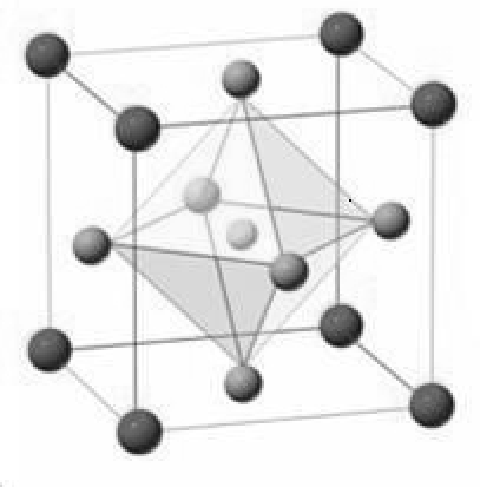}
&
\includegraphics[width=0.4\textwidth]{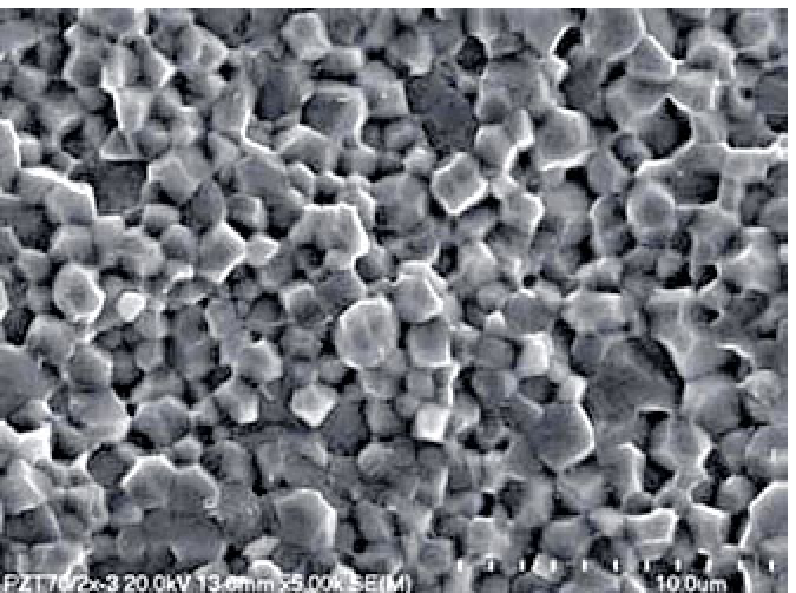}
\\
\includegraphics[width=0.4\textwidth]{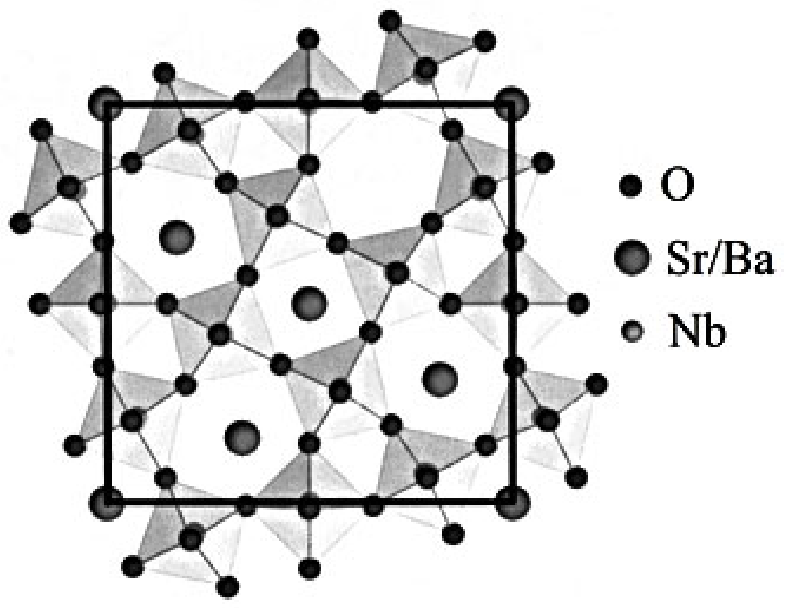}
&
\includegraphics[width=0.4\textwidth]{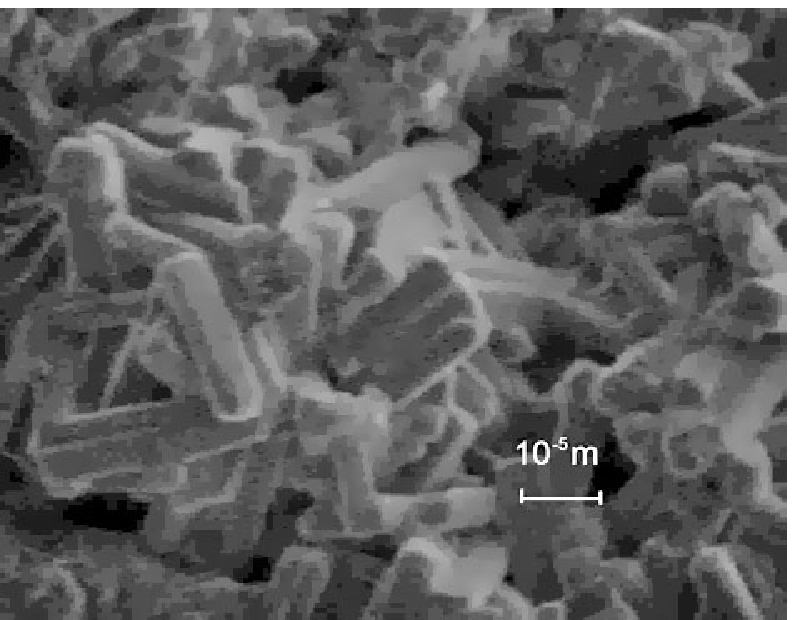}
\end{tabular}
\end{center}
\caption{Ceramics with symmetric (a, b) and asymmetric (c, d) unit cells:
(a) perovskite type structure, (b) SEM image of PZT ceramics with perovskite type structure, (c) tetragonal tungsten bronze type structure, (d) SEM image of Sr$_{0.7}$Ba$_{0.3}$Nb$_2$O$_6$ ceramics with tetragonal tungsten bronze type structure.
}
\label{fig1}
\end{figure}

Figure~\ref{fig1} shows SEM micrographs of ceramics with symmetric [perovskite, figure~\ref{fig1}~(a)] and asymmetric [tetragonal tungsten bronze type structure, figure~\ref{fig1}~(c)]. The presented ceramics was obtained from powders synthesized using the sol-gel method. Figure~\ref{fig1}~(d) clearly shows that ceramics with an asymmetric unit cell grow anisotropically. There were obtained textured ceramics-polymer composites from such ceramics earlier \cite{5}.
	
There was selected ceramic bismuth titanate (Bi$_4$Ti$_3$O$_{12}$) which is a ferroelectric material with layered structure belonging to the Aurivillius family. Its crystalline structure consists of (Bi$_2$Ti$_3$O$_{10}$)$^{2-}$ layers formed by BiTiO$_3$ unit cells of perovskite-like structures alternating with (Bi$_2$O$_2$)$^{2+}$ layers perpendicular to the $c$ axis.  Bi$_4$Ti$_3$O$_{12}$ has a strong potential for being employed in many electronic devices because of its relatively high Curie temperature $T_\mathrm{c}=\SI{675}{\degreeCelsius}$, low dielectric loss and high anisotropy \cite{6}.

Particle size of starting materials is a very important factor for the grain orientation of annealed ceramics. Highly textured ceramics can only be obtained by using nanosized powders \cite{7}. The sol-gel method was used in this work to obtain ceramic powders, which also makes it possible to get nanopowders.

\section{Obtaining bismuth titanate powders by sol-gel method}

Bismuth nitrate pentahydrate Bi(NO$_3$)$_3$\,$\cdot$\,5H$_2$O (POCH) and tetrabutyl orthotitanate Ti(CH$_3$(CH$_2$)$_3$O)$_4$ (Fluka) were used as precursors. Bismuth nitrate pentahydrate was dissolved in 2-methoxyethanol CH$_3$OCH$_2$CH$_2$OH (Riedel-de Ha\"{e}n) at \SI{40}{\degreeCelsius}. Separately, tetrabutyl orthotitanate was dissolved in 2-methoxyethano at room temperature. Acetylacetone CH$_3$COCH$_2$COCH$_3$ (Merck) was subsequently added into this mixture as the chelating agent. The titanium solution was then added into the bismuth solution and mixed together for 2 h.

\begin{figure}[!tb]
\centerline{
\includegraphics[width=0.48\textwidth]{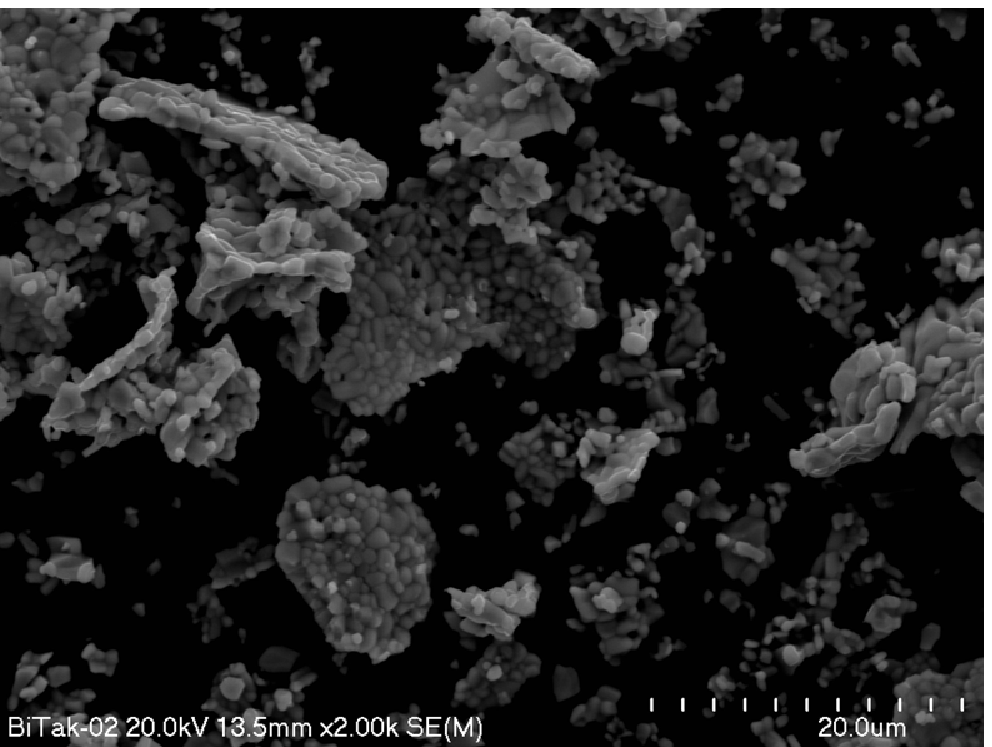}
\hspace{2mm}
\includegraphics[width=0.48\textwidth]{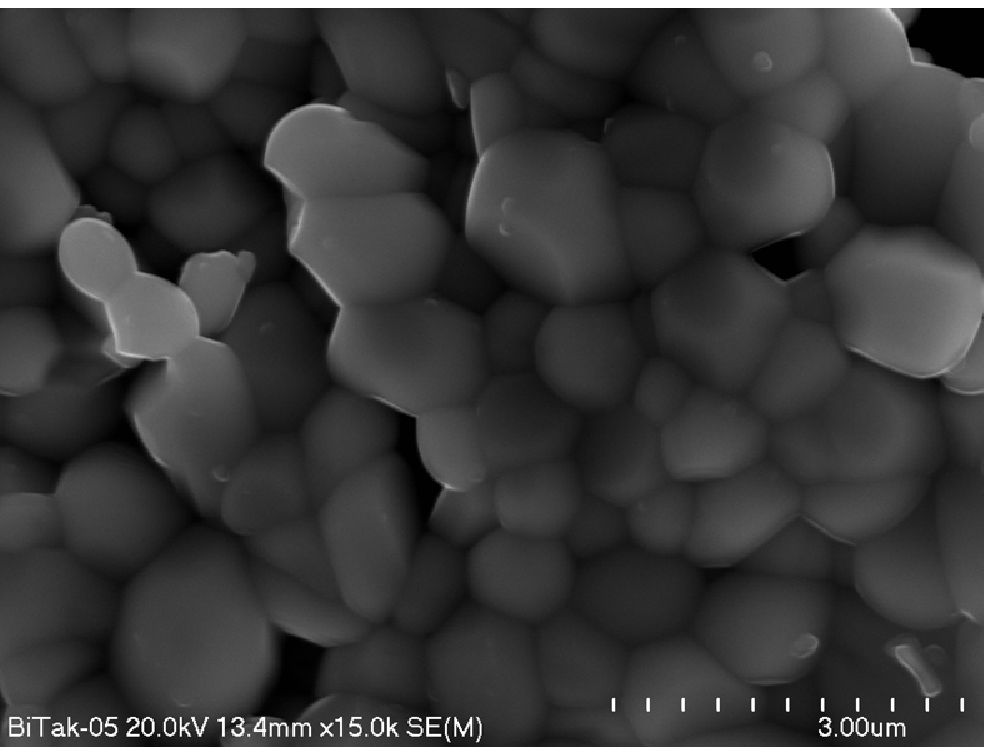}
}
\caption{SEM micrographs of Bi$_4$Ti$_3$O$_{12}$ powders synthesized by sol-gel method (at different magnifications).}
\label{fig2}
\end{figure}

The proposed equation for the formation of the Bi$_4$Ti$_3$O$_{12}$ phase is shown in equation (\ref{eq:1}).
\begin{eqnarray}
\label{eq:1}
4\text{Bi}(\text{NO}_3)_3 \cdot 5\text{H}_2\text{O} + 3\text{Ti}(\text{CH}_3(\text{CH}_2)_3\text{O})_4 + 72\text{O}_2 \rightarrow \text{Bi}_4\text{Ti}_3\text{O}_{12} + 6\text{N}_2\text{O}_5\uparrow +48\text{CO}_2\uparrow + 59\text{H}_2\text{O}\uparrow \, .
\end{eqnarray}
Then, distilled water was added to initiate hydrolysis and gel formation. The gel obtained was dried at room temperature, and then heated at \SI{850}{\degreeCelsius} for 3 hours in order to eliminate organic parts. The powder obtained after calcination was ground. The  morphology of powders was examined using a scanning electron microscope HITACHI S-4700 with microanalysis system NORAN Vantage. Qualitative and quantitative analysis of the chemical composition (EDS) was carried out. It was confirmed that the chemical composition of the powders obtained is in conformity with the previous assumptions. The SEM micrographs of Bi$_4$Ti$_3$O$_{12}$ powders are shown in figure~\ref{fig2}. The grains in figure~\ref{fig2} exceed 500~nm; they are very large compared to other ceramic powders obtained using the sol-gel method. Using the sol-gel method, the co-author of this article previously obtained many ceramic materials, such as: PZT, PLZT, BaTiO$_3$, SBN \cite{8} and always obtained powders with nanometer grains.

\begin{figure}[!b]
\centerline{
\includegraphics[width=0.48\textwidth]{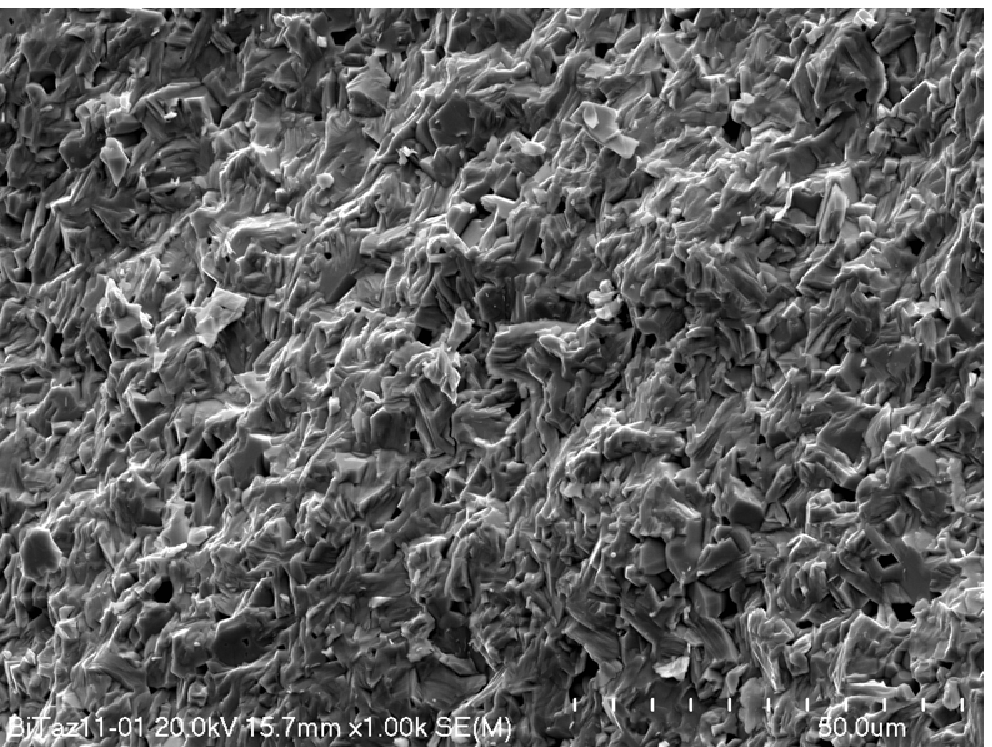}
\hspace{2mm}
\includegraphics[width=0.48\textwidth]{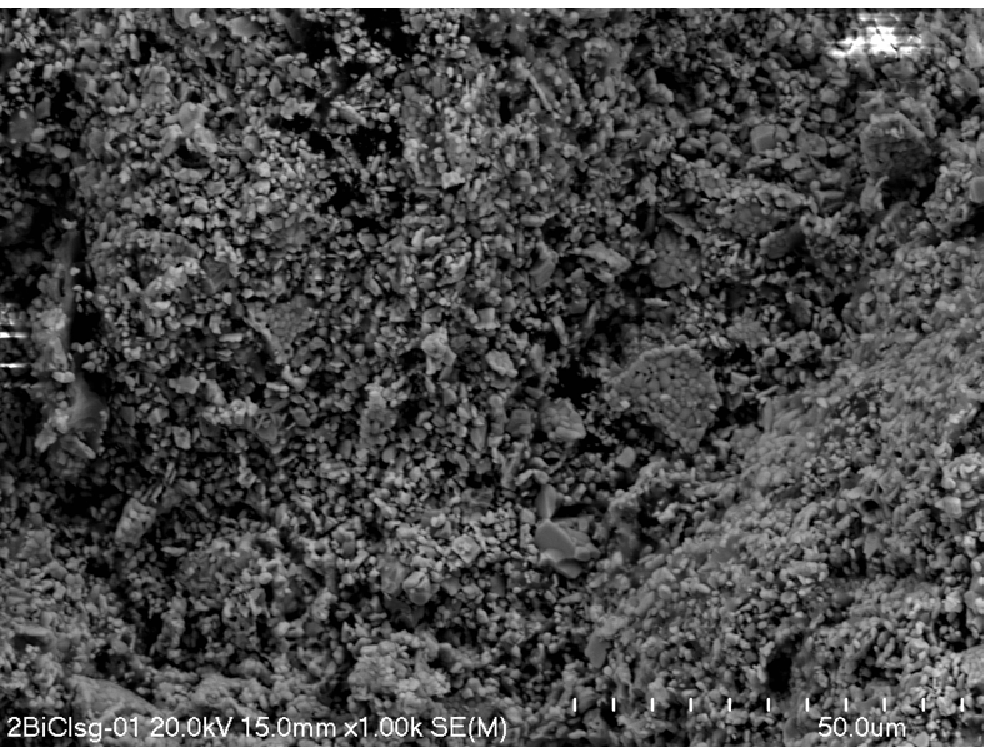}
}
\parbox[t]{0.48\textwidth}{%
\caption{SEM micrograph of Bi$_4$Ti$_3$O$_{12}$ ceramic sintered at \SI{1100}{\degreeCelsius}.}\label{fig3}
}%
\parbox[t]{0.48\textwidth}{%
\caption{SEM micrograph of Bi$_4$Ti$_3$O$_{12}$-PVC composite with volume fraction of ceramics $\Phi=0.2$.}
\label{fig4}
}
\end{figure}

The powders were used to fabricate a bulk ceramics. The SEM micrograph of Bi$_4$Ti$_3$O$_{12}$ ceramics sintered at \SI{1100}{\degreeCelsius} is shown in figure~\ref{fig3}.

\section{Bi$_4$Ti$_3$O$_{12}$-PVC ceramics-polymer composites}

The Bi$_4$Ti$_3$O$_{12}$ powder was used to fabricate a composite with the volume fraction of ceramics $\Phi=0.2$. The ceramic powders were placed in a polyvinyl chloride (PVC) polymer matrix. A hot pressing method was used to make composites of 0-3 connectivity (a polymer phase three dimensionally connected with isolated ceramic particles). SEM micrograph of the composite is shown in figure~\ref{fig4}. Unfortunately, the obtained composite is not textured.

\section{Summary}

The purpose of this work was to obtain Bi$_4$Ti$_3$O$_{12}$-PVC textured ceramics-polymer composites. Despite a proper (according to the authors) selection of a test material and a suitable method for its obtaining, the attempt to arrive at a composite with an oriented structure failed. Nevertheless, the authors consider the results of their research to be quite useful since they show that nanopowders should be used for obtaining textured composites. In this case, the reason behind the failure was the excessive size of the ceramic powder grains. It should be mentioned that Bi$_4$Ti$_3$O$_{12}$ powders obtained by other authors using wet chemical methods are not nanometer as well, e.g.: \cite{4,9}.

\ukrainianpart

\title{Спроба отримання  Bi$_4$Ti$_3$O$_{12}$-PVC текстурованих
керамічно-полімерних композитів}

\author{Е. Ногас-Цвікель\refaddr{label1}, Г. Бернар\refaddr{label2}}
\addresses{
\addr{label1} Факультет комп'ютерних наук і матеріалознавства,
Університет Сілезії, м. Сосновєц, Польща \addr{label2} Стипендіат
проекту ``DoktoRIS'', Польща }

 \makeukrtitle
\begin{abstract}

Виготовлено композити Bi$_4$Ti$_3$O$_{12}$-PVC. Керамічні порошки
титанату вісмута приготовані золь-гель методом, використовуючи
пентагідрат нітрату вісмута Bi(NO$_3$)$_3$ $\cdot$ 5 H$_2$O і
тетрабутил титанат  Ti(CH$_3$(CH$_2$)$_3$O)$_4$ в якості
прекурсорів. Композити Bi$_4$Ti$_3$O$_{12}$-PVC виготовлені з
керамічних порошків і полімерних порошків методом гарячого
пресування.

\keywords керамічно-полімерні композити, золь-гель, титанат вісмута,
Bi$_4$Ti$_3$O$_{12}$
\end{abstract}

\end{document}